# A SCALABLE, LEXICON BASED TECHNIQUE FOR SENTIMENT ANALYSIS

Chetan Kaushik and Atul Mishra

Computer Engg. Department, YMCA University of Science & Technology, Faridabad

## ABSTRACT

*Rapid increase in the volume of sentiment rich social media on the web has resulted in an increased interest among researchers regarding Sentimental Analysis and opinion mining. However, with so much social media available on the web, sentiment analysis is now considered as a big data task. Hence the conventional sentiment analysis approaches fails to efficiently handle the vast amount of sentiment data available now a days. The main focus of the research was to find such a technique that can efficiently perform sentiment analysis on big data sets. A technique that can categorize the text as positive, negative and neutral in a fast and accurate manner. In the research, sentiment analysis was performed on a large data set of tweets using Hadoop and the performance of the technique was measured in form of speed and accuracy. The experimental results shows that the technique exhibits very good efficiency in handling big sentiment data sets.*

## 1. INTRODUCTION

Sentiment Analysis is a Natural Language Processing task that includes obtaining the writer's feeling about several products or services which are posted on the internet via various posts or comments. In other words Sentiment Analysis helps in determining the response of a user or a group of users on a topic and categorizing their opinion as positive, negative or neutral. In recent years, a huge increase is seen in the amount of opinions expressed on the web by social media users and hence it has captured the interest of more and more researchers.

Liu [1] defined a sentiment as a quintuple "<oj, fjk, soijkl, hi, tl >, where oj is a target object, fjk is a feature of the object oj, soijkl is the sentiment value of the opinion of the opinion holder hi on feature fjk of object oj at time tl, soijkl is +ve, -ve, or neutral, or a more granular rating, hi is an opinion holder, tl is the time when the opinion is expressed."

Sentiment Analysis has its applications in several fields such as in marketing for knowing customers' response towards a new product or service or by social media users to find the general opinion about a topic which is currently trending. It can help manufactures in deciding the strategies for the launch of new products based on the responses towards the previous versions of that product in various geographical areas. It can also be used to detect heated arguments in comments, use of abusive language or spam detection.

Sentimental Analysis can be phrase based in which the sentiment of a single phrase is taken in to consideration, sentence based in which the sentiment of the whole sentence is calculated or it can be document based in which an aggregated sentiment of the whole document is categorized as positive, negative or neutral

Sentiment Analysis is generally carried out in three steps. First, the subject towards which the sentiment is directed is found then, the polarity of the sentiment is calculated and finally the degree of the polarity is assigned with the help of a sentiment score which denotes the intensity of the sentiment.

Over the last few years, the amount of sentiment rich data on the web has increased rapidly with the increase in social media users. Various companies are now turning to online forums, blogs etc. to get the review of their products from customers. The sentiment data hence is now considered as big data which needs techniques that are capable of handling such large amounts of data efficiently.

In this research a lexicon based approach is used to perform sentiment analysis. There has been a lot of research on sentiment calculation using lexicon based and other techniques. Some of which are described in the following section.

## 2. RELATED WORK

A lot of work has been done in the field of opinion mining or sentiment analysis for well over a decade now. Different techniques are used to classify the text according to polarity. Most of these techniques can be classified under two categories:

### 2.1. Machine learning

Machine learning strategies work by training our algorithm with a training data set before applying it to the actual data set. That is, a machine learning algorithm needs to be trained first for both supervised and unsupervised learning tasks.

Machine learning techniques first trains the algorithm with some particular inputs with known outputs so that later it can work with new unknown data. Some of the most renowned works based on machine learning are as follows:

Melville [2], Rui [3], Ziqiong [4], Songho [5], Qiang [6] and Smeureanu [7] used naïve bayes for classification of text. Naïve bayes is one of the most popular method in text classification. It is considered as one of the most simple and efficient approaches in NLP. It works by calculating the probability of an element being in a category. First the prior probability is calculated which is afterwards multiplied with the likelihood to calculate the final probability. The method assumes every word in the sentence to be independent which makes it easier to implement but less accurate. That's why the method is given the name 'naïve'.

Rui [3], Ziqiong [4], Songho [5], and Rudy [8] used SVM (Support Vector Machines) in their approach. The sentiment classification was done using discriminative classifier. This approach is based on structural risk minimization in which support vectors are used to classify the training data sets into two different classes based on a predefined criteria. Multiclass SVM can also be used for text classification [9].

Songho [5] used centroid classifier which assign a centroid vector to different training classes and then uses this vector to calculate the similarity values of each element with a class. If the similarity value exceeds a defined threshold then the element is assigned to that class (polarity in this case).

Songho [5] also used K-Nearest Neighbor (KNN) approach which finds 'K' nearest neighbours of a text document among the documents in the training data set. Then classification is performed on the basis of similarity score of a class with respect to a neighbour.

Some approaches use prediction method. Winnow first predicts a class for an entity and then feedback is used to check for mistakes. The weight vectors are changed according to the feedback and the process is repeated over a large training data set.

A combined approach was proposed based on included rule classification, machine learning and supervised learning [8]. Each training set was subjected to a 10 fold cross validation process. Several classifiers were used in series. Whenever a classifier failed, the text was passed on to the next classifier until all the texts are classified or there are no remaining classifiers Rui [3] used Ensemble technique which combines the output of several classification methods into a single integrated output.

Another approach based on artificial neural networks was used to divide the document into positive, negative and fuzzy tone [10]. The approach was based on recursive least squares back propagation training algorithm.

Long-Sheng [11] used a neural network based approach to combine the advantages of machine learning and information retrieval techniques.

### 2.2. Lexicon Based

Lexicon Based techniques work on an assumption that the collective polarity of a document or sentence is the sum of polarities of the individual words or phrases. Some of the significant works done using this technique are:

Kamps [12] used a simple technique based on lexical relations to perform classification of text. Andrea [13] used word net to classify the text using an assumption that words with similar polarity have similar orientation.

Ting-Chun [14] used an algorithm based on pos (part of speech) patter. A text phrase was used as a query for a search engine and the results were used to classify the text.

Prabhu [15] which used a simple lexicon based technique to extract sentiments from twitter data Turney [19] used semantic orientation on user reviews to identify the underlying sentiments. Taboada [20] used lexicon based approach to extract sentiments from microblogs.

Sentiment analysis for microblogs is more challenging because of problems like use of short length status message, informal words, word shortening, spelling variation and emoticons. Twitter data was used for sentiment analysis by [21].

Negation word can reverse the polarity of any sentence. Taboada [20] performed sentiment analysis while handling negation and intensifying words. Role of negation was surveyed by [25]. Minquing [26] classified the text using a simple lexicon based approach with feature detection.
It was observed that most of these existing techniques doesn't scale to big data sets efficiently. While various machine learning methodologies exhibits better accuracy than lexicon based techniques, they take more time in training the algorithm and hence are not suitable for big data sets. In this paper, lexicon based approach is used to classify the text according to polarity.

# 3. PROPOSED APPROACH

As explained earlier the focus of this research was to device an approach that can perform sentiment analysis quicker because vast amount of data needed to be analyzed. Also, it had to be made sure that accuracy is not compromised too much while focusing on speed.

The proposed approach is a lexicon based technique i.e. a dictionary of sentiment bearing words along with their polarities was used to classify the text into positive, negative or neutral opinion. Machine learning techniques are not used because although they are more accurate than the lexicon based approaches, they take far too much time performing sentiment analysis as they have to be trained first and hence are not efficient in handling big sentiment data.

The main component of this approach is the sentiment lexicon or dictionary.

## 3.1. Sentiment Lexicon

Unlike the small sized dictionaries which are used in other approaches, which are build up from a small set of words using a training data set, the dictionary used in this technique contains a large set of sentiment bearing words along with their polarity.

The dictionary is domain specific i.e. the polarities of the words in the dictionary are set according to a specific domain e.g. book reviews, political blogs etc. Same word in different domains can have different meanings, the dictionary used in this approach is made for movie review domain.

The dictionary contains all forms of a word i.e. every word is stored along with its various verb forms e.g. applause, applauding, applauded, applauds. Hence eliminating the need for stemming each word which saves more time.

Emoticons are generally used by people to depict emotions. Hence it is obvious that they contain very useful sentiment information in them. The dictionary used in the approach contains more than 30 different emoticons along with their polarities.

The Dictionary also contains the strength of the polarity of every word. Some word depicts stronger emotions than others. For example good and great are both positive words but great depicts a much stronger emotion.

Negation and blind negation are very important in identifying the sentiments, as their presence can reverse the polarity of the sentence. The dictionary used here also contains various negation and blind negation words so that they can be identified in the sentence.

The dictionary contains several different adjectives, nouns, negation words, emoticons etc. Some of these words are shown as an example in table 1.

Table 1. Some examples of the words and emoticons in the sentiment dictionary

| Strength | Word | POS | Stemmed | Polarity |
| --- | --- | --- | --- | --- |
| weaksubj | abandoned | adj | n | negative |
| weaksubj | abandonment | noun | n | negative |
| weaksubj | abandon | verb | y | negative |
| strongsubj | needed | verb | n | blindnegation |

| strongsubj | require | verb  | n | blindnegation |
|------------|---------|-------|---|---------------|
| strongsubj | not     | advb  | n | negation      |
| strongsubj | neither | conj  | n | negation      |
| strongsubj | nor     | conj  | n | negation      |
| strongsubj | :)      | emoti | n | positive      |
| strongsubj | :(      | emoti | n | negative      |

### 3.2. Feature detection using Twitter hashtags

A hashtag is a word or an unspaced phrase prefixed with the number sign ("#"). It is a form of metadata tag. Words in messages on microblogging and social networking services such as Twitter, Facebook, Google+ or Instagram may be tagged by putting "#" before them, either as they appear in a sentence or they are appended to it.

In this research, twitter data is categorized according to polarity. Detecting the subject towards which the sentiment is directed is a tedious task to perform, but as twitter is used as data source hashtags can be used to easily identify the subject hence eliminating the need for using a complex mechanism for feature detection thereby saving time and effort.

### 3.3. Handling negation and blind negation

Negation words are the words which reverse the polarity of the sentiment involved in the text. For example 'the movie was not good'. Although the word 'good' depicts a positive sentiment the negation – 'not' reverses its polarity. In the proposed approach whenever a negation word is encountered in a tweet, its polarity is reversed.

Blind negation words are the words which operates on the sentence level and points out a feature that is desired in a product or service. For example in the sentence 'the acting needed to be better', 'better' depicts a positive sentiment but the presence of the blind negation word- 'needed' suggests that this sentence is actually depicting negative sentiment. In the proposed approach whenever a blind negation word occurs in a sentence its polarity is immediately labelled as negative.

### 3.4. Sentiment Calculation

Sentiment calculation is done for every tweet and a polarity score is given to it. If the score is greater than 0 then it is considered to be positive sentiment on behalf of the user, if less than 0 then negative else neutral. The polarity score is calculated by using algorithm 1.

Algorithm 1: ALGO_SENTICAL

Input: Tweets, SentiWord_Dictionary Output: Sentiment (positive, negative or neutral)
   1) BEGIN
   2) For each tweet T$i$
   3) {
   4)   SentiScore = 0;
   5)   For each word W$j$ in T$i$ that exists in Sentiword_Dictionary
   6)   {
   7)     If polarity[W$j$] = blindnegation

```
8)       {
9)          Return negative;
10)      }
11)   Else
12)   {
13)       If polarity[Wj] = positive && strength[Wj] = Strongsubj
14)       {
15)        SentiScore = SentiScore + 1;
16)       }
17)       Else If polarity[Wj] = positive && strength[Wj] = Weaksubj
18)       {
19)          SentiScore = SentiScore + 0.5;
20)       }
21)       Else If polarity[Wj] = negative && strength[Wj] = Strongsubj
22)       {
23)           SentiScore = SentiScore – 1;
24)       }
25)       Else If polarity[Wj] = negative && strength[Wj] = Weaksubj
26)       {
27)       SentiScore = SentiScore – 0.5;
28)       }
29)    }
30)   If polarity[Wj] = negation
31)   {
32)   Sentiscore = Sentiscore * -1
33)   }
34)  }
35)   If Sentiscore of Ti >0
36)   {
37)  Sentiment = positive
38)   }
39)  Else If Sentiscore of TI<0
40)   {
41)   Sentiment = negative
42)   }
43)   Else
44)   {
45)    Sentiment = neutral
46)    }
47) Return Sentiment
48)  }
49) END
```

## 4. PERFORMANCE EVALUATION

In this section, description of the tools which were used to perform sentiment analysis is given along with the performance of the approach and its comparison with existing approaches.

### 4.1. Experimental Setup

The proposed algorithm was implemented using a sandboxed version of Hadoop. Hadoop is designed to work in a huge multimode environment consisting of thousands of servers situated at

International Journal in Foundations of Computer Science & Technology (IJFCST), Vol.4, No.5, September 2014

different locations. But for research purposes often a single node virtual environment is used that creates an illusion of several nodes which are situated at different locations and are working together. An Intel Core i5-3210M CPU@2.50GHz processor with 6 GB memory was used to simulate the Hadoop Environment. Data was imported from Twitter using Flume, a distributed, reliable, and available service for efficiently collecting, aggregating, and moving large amounts of streaming data into the Hadoop Distributed File System (HDFS).

Then the data is refined and the algorithm described in the previous section is implemented upon it with the help of HiveQL. HiveQL is an SQL like language which is used in hadoop for interacting and manipulating huge databases.

### 4.2. Results and Evaluation

As explained earlier the purpose of this research was to device a method that can quickly compute the sentiments of huge data sets without compromising too much with accuracy.

The proposed approach has performed very well in terms of speed. It took 14.8 Seconds to analyse the sentiments behind 6,74,412 tweets. Table 1 shows the performance of the algorithm. Moreover the speed can be substantially improved by using an actual multimode hadoop environment instead of a single node sandboxed configuration.

The algorithm also performed very well in terms of accuracy. Tweets were analyzed with an accuracy of 73.5 %. As expected it is not as high as the machine learning methodologies but it shows a fairly good accuracy when compared with other lexicon based approaches. Figure 1 shows a comparison between average accuracies of machine learning and lexicon based techniques discussed in section II and the accuracy of the proposed approach.

Table 2 Performance of the proposed approach

| Number of Tweets | Time Taken | Accuracy |
|---|---|---|
| 6,74,412 | 14.8 seconds | 73.5 % |

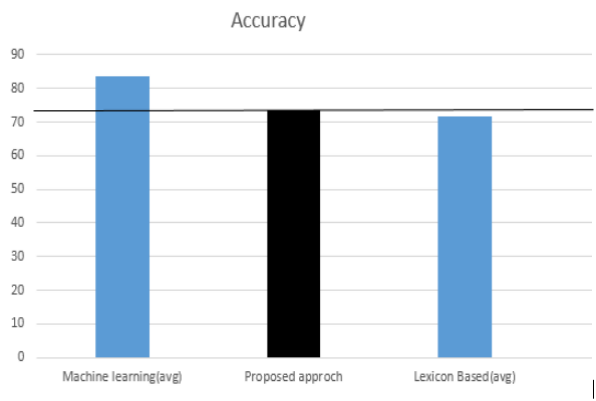

Figure 1. Comparison of the proposed approach with other methodologies

## 5. CONCLUSION AND FUTURE WORK

Sentiment analysis is being used for different applications and can be used for several others in future. It is evident that its applications will definitely expand to more areas and will continue to encourage more and more researches in the field. We have done an overview of some state-of-the-art solutions applied to sentiment classification and provided a new approach that scales to big data sets efficiently.

A scalable and practical lexicon based approach for extracting sentiments using emoticons and hashtags is introduced. Hadoop was used to classify Twitter data without need for any kind of training.

Our approach performed extremely well in terms of both speed and accuracy while showing signs that it can be further scaled to much bigger data sets with similar, in fact better performance.
In this research, main focus was on performing sentiment analysis quickly so that big data sets can be handled efficiently. The work can be further expanded by introducing techniques that increase the accuracy by tackling problems like thwarted expressions and implicit sentiments which still needs to be resolved properly.

Also as explained earlier, this work was implemented on a single node sandboxed configuration and although it is expected that it will perform much better in a multimode enterprise level configuration, it is desirable to check its performance in such environment in future.

## REFERENCES

[1] B. Liu. Sentiment Analysis and Subjectivity. Handbook of Natural Language Processing, Second Edition, (editors: N. Indurkhya and F. J. Damerau), 2010.
[2] Melville, Wojciech Gryc, "Sentiment Analysis of Blogs by Combining Lexical Knowledge with Text Classification", KDD"09, June 28–July 1, 2009, Paris, France.Copyright 2009 ACM 978-1-60558-495-9/09/06.
[3] Rui Xia, Chengqing Zong, Shoushan Li, "Ensemble of feature sets and classification algorithms for sentiment classification", Information Sciences 181 (2011) 1138–1152.
[4] Ziqiong Zhang, Qiang Ye, Zili Zhang, Yijun Li, "Sentiment classification of Internet restaurant reviews written in Cantonese", Expert Systems with Applications xxx (2011) xxx–xxx.
[5] Songbo Tan, Jin Zhang, "An empirical study of sentiment analysis for chinese documents", Expert Systems with Applications 34 (2008) 2622–2629.
[6] Qiang Ye, Ziqiong Zhang, Rob Law, "Sentiment classification of online reviews to travel destinations by supervised machine learning approaches", Expert Systems with Applications 36 (2009) 6527–6535.
[7] Ion SMEUREANU, Cristian BUCUR, "Applying Supervised Opinion Mining Techniques on Online User Reviews", Informatica Economică vol. 16, no. 2/2012.
[8] Rudy Prabowo, Mike Thelwall, "Sentiment analysis: A combined approach." Journal of Informetrics 3 (2009) 143–157.
[9] Kaiquan Xu , Stephen Shaoyi Liao , Jiexun Li, Yuxia Song, "Mining comparative opinions from customer reviews for Competitive Intelligence", Decision Support Systems 50 (2011) 743–754.
[10] ZHU Jian , XU Chen, WANG Han-shi, "" Sentiment classification using the theory of ANNs", The Journal of China Universities of Posts and Telecommunications, July 2010, 17(Suppl.): 58–62 .[16] Ziqiong Zhang, Qiang Ye, Zili Zhang, Yijun Li, "Sentiment classification of Internet restaurant reviews written in Cantonese", Expert Systems with Applications xxx (2011)
[11] Long-Sheng Chen, Cheng-Hsiang Liu, Hui-Ju Chiu, "A neural network based approach for sentiment classification in the blogosphere", Journal of Informetrics 5 (2011) 313–322.

[12] Kamps, Maarten Marx, Robert J. Mokken and Maarten De Rijke, "Using wordnet to measure semantic orientation of adjectives", Proceedings of 4th International Conference on Language Resources and Evaluation, pp. 1115-1118, Lisbon, Portugal, 2004.
[13] Andrea Esuli and Fabrizio Sebastiani, "Determining the semantic orientation of terms through gloss classification", Proceedings of 14th ACM International Conference on Information and Knowledge Management,pp. 617-624, Bremen, Germany, 2005.
[14] Ting-Chun Peng and Chia-Chun Shih , "An Unsupervised Snippet-based Sentiment Classification Method for Chinese Unknown Phrases without using Reference Word Pairs", 2010 IEEE/WIC/ACM International Conference on Web Intelligence and intelligent Agent Technology JOURNAL OF COMPUTING, VOLUME 2, ISSUE 8, AUGUST 2010, ISSN 2151-9617 .
[15] Prabu Palanisamy, Vineet Yadav, Harsha Elchuri, "Serendio: Simple and Practical lexicon based approach to Sentiment Analysis", Serendio Software Pvt Ltd, 2013.
[16] Subhabrata Mukherjee, Dr. Pushpak Bhattacharyya, Indian Institute of Technology, Bombay, Department of Computer Science and Engineering, June 2012.
[17] Bickel, S.Bruckner, M. Scheffer, "Discriminative learning for differing training and test distributions", International Conference on Machine Learning, 2007.
[18] Benamara, Farah, Carmine Cesarano, Antonio Picariello, Diego Reforgiato and VS Subrahmanian, "Sentiment analysis: Adjectives and adverbs are better than adjectives alone" International Conference on Weblogs and Social Media, ICWSM, Boulder, CO. 2007.
[19] Peter Turney and Michael Littman. 2003. Measuring praise and criticism: Inference of semantic orientationfrom association. ACM Transactions on Information Systems 21(4):315–346.
[20] Maite Taboada, Julian Brooke, Milan Tofiloski, Kimberly Voll and Manfred Stede. 2011. Lexicon-based methods for sentiment analysis. Computational linguistics, volume 37, number2, 267–307, MIT Press.
[21] Albert Bifet and Eibe Frank. 2010. Sentiment knowledge discovery in twitter streaming data, DiscoveryScience 1–14, Springer.
[22] Bo Han and Timothy Baldwin. 2011. Lexical normalisation of short text messages: Makn sens a# twitter.Proceedings of the 49th Annual Meeting of the Association for Computational Linguistics: Human Language Technologies Volume 1 , 368–378.
[23] Max Kaufmann and Jugal Kalita. 2010. Syntactic normalization of Twitter messages. International Conference on Natural Language Processing Kharagpur, India.
[24] Dmitry Davidov, Oren Tsur and Ari Rappoport. 2010. Enhanced sentiment learning using twitter hashtagsand smileys. Proceedings of the 23rd International Conference on Computational Linguistics 241 –249, Association for Computational Linguistics.
[25] Michael Wiegand, Alexandra Balahur, Benjamin Roth, Dietrich Klakow, Andŕes Montoyo. 2010. A survey on the role of negation in sentiment analysis. Proceedings of the workshop on negation and speculation in natural language processing 60–68, Association for Computational Linguistics.
[26] Minqing Hu, Bing Liu. Mining and Summarizing Customer Reviews, Department of Computer Science, University of Illinois at Chicago, Research Track Paper.

## Authors

**Chetan Kaushik**[1] is a student in the Department of Computer Engineering, YMCA University of Science and Technology, Haryana, India. He holds a bachelor's degree in information Technology from Kurukshetra University, kurukshetra and is pursuing his master's degree in computer engineering from the Department of Computer Engineering, YMCA University of Science and Technology. His research interests includes big data and Social network mining.

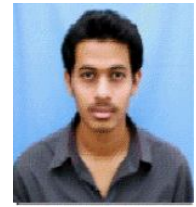

**Atul Mishra**[2] is working as an Associate Professor in the Department of Computer Engineering, YMCA University of Science and Technology, Haryana, India. He holds a Master's Degree in Computer Science and Technology, from University of Roorkee(now IIT,Roorkee) and PhD in Computer Science and Engg. from MD University, Rohtak. He has about 16 years of work experience in the Optical Telecommunication Industry specializing in Optical Network Planning and Network Management tools development. His research interests include SOA, Network Management & Mobile Agents.

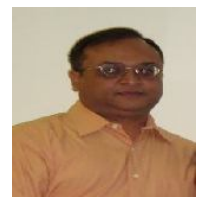